\documentstyle[12pt,epsf]{article}
\expandafter\ifx\csname amssym12.def\endcsname\relax \else\endinput\fi
\expandafter\edef\csname amssym12.def\endcsname{%
       \catcode`\noexpand\@=\the\catcode`\@\space}
\catcode`\@=11
\def\undefine#1{\let#1\undefined}
\def\newsymbol#1#2#3#4#5{\let\next@\relax
 \ifnum#2=\@ne\let\next@\msafam@\else
 \ifnum#2=\tw@\let\next@\msbfam@\fi\fi
 \mathchardef#1="#3\next@#4#5}
\def\mathhexbox@#1#2#3{\relax
 \ifmmode\mathpalette{}{\m@th\mathchar"#1#2#3}%
 \else\leavevmode\hbox{$\m@th\mathchar"#1#2#3$}\fi}
\def\hexnumber@#1{\ifcase#1 0\or 1\or 2\or 3\or 4\or 5\or 6\or 7\or 8\or
 9\or A\or B\or C\or D\or E\or F\fi}
\font\tenmsa=msam10 scaled\magstep1
\font\sevenmsa=msam7 scaled\magstep1
\font\fivemsa=msam5 scaled\magstep1
\newfam\msafam
\textfont\msafam=\tenmsa
\scriptfont\msafam=\sevenmsa
\scriptscriptfont\msafam=\fivemsa
\edef\msafam@{\hexnumber@\msafam}
\mathchardef\dabar@"0\msafam@39
\def\dashrightarrow{\mathrel{\dabar@\dabar@\mathchar"0\msafam@4B}}
\def\dashleftarrow{\mathrel{\mathchar"0\msafam@4C\dabar@\dabar@}}

\def\ulcorner{\delimiter"4\msafam@70\msafam@70 }
\def\urcorner{\delimiter"5\msafam@71\msafam@71 }
\def\llcorner{\delimiter"4\msafam@78\msafam@78 }
\def\lrcorner{\delimiter"5\msafam@79\msafam@79 }
\def\yen{{\mathhexbox@\msafam@55 }}
\def\checkmark{{\mathhexbox@\msafam@58 }}
\def\circledR{{\mathhexbox@\msafam@72 }}
\def\maltese{{\mathhexbox@\msafam@7A }}
\font\tenmsb=msbm10 scaled\magstep1
\font\sevenmsb=msbm7 scaled\magstep1
\font\fivemsb=msbm5 scaled\magstep1
\newfam\msbfam
\textfont\msbfam=\tenmsb
\scriptfont\msbfam=\sevenmsb
\scriptscriptfont\msbfam=\fivemsb
\edef\msbfam@{\hexnumber@\msbfam}

\def\widehat#1{\setbox\z@\hbox{$\m@th#1$}%
 \ifdim\wd\z@>\tw@ em\mathaccent"0\msbfam@5B{#1}%
 \else\mathaccent"0362{#1}\fi}
\def\widetilde#1{\setbox\z@\hbox{$\m@th#1$}%
 \ifdim\wd\z@>\tw@ em\mathaccent"0\msbfam@5D{#1}%
 \else\mathaccent"0365{#1}\fi}
\font\teneufm=eufm10 scaled\magstep1
\font\seveneufm=eufm7 scaled\magstep1
\font\fiveeufm=eufm5 scaled\magstep1
\newfam\eufmfam
\textfont\eufmfam=\teneufm
\scriptfont\eufmfam=\seveneufm
\scriptscriptfont\eufmfam=\fiveeufm

\csname amssym.def\endcsname
\newif{\ifcomentarios}
\comentariosfalse
\renewcommand{\theequation}{\thesection.\arabic{equation}}

\newcommand{\be}{\begin{equation}}
\newcommand{\ee}{\end{equation}}
\newcommand{\bma}{\begin{displaymath}}
\newcommand{\ema}{\end{displaymath}}
\newcommand{\bc}{\begin{center}}
\newcommand{\ec}{\end{center}}
\newcommand{\text}{\rm}

\newcommand{\uflex}
{{\scriptstyle {\raise 9pt\hbox{$\backslash$}\,\!\!\!\!\!\Bigg\vert}}}
\newcommand{\ncm}{\newcommand}

\ncm{\rncm}{\renewcommand}
\ncm{\id}{{\bf 1}}
\ncm{\beq}{\begin{equation}}
\ncm{\eeq}{\end{equation}}
\ncm{\ba}{\begin{array}}
\ncm{\bea}{\begin{eqnarray}}
\ncm{\beanon}{\begin{eqnarray*}}
\ncm{\ea}{\end{array}}
\ncm{\eea}{\end{eqnarray}}
\ncm{\eeanon}{\end{eqnarray*}}
\ncm{\fns}{\footnotesize}
\ncm{\setc}[1]{\setcounter{equation}{#1}}
\newcounter{eqnr}
\newenvironment{eqnarrayabc}{\stepcounter{equation}
  \setcounter{eqnr}{\value{equation}}\setc{0}
  \rncm{\theequation}{\thesection.\arabic{eqnr}\alph{equation}}
  \begin{eqnarray}}{\end{eqnarray}\setc{\value{eqnr}}}
\ncm{\eqboxabc}[3]{\newline\parbox[t]{1.5cm}{#1}\hfill
  \parbox[b]{12cm}{\begin{eqnarray*} #3\end{eqnarray*}}\hfill
   \parbox[b]{1.5cm}{\vspace{-0.0cm}\begin{eqnarrayabc}#2\end{eqnarrayabc}}\newline}
\ncm{\eqbox}[2]{\newline\parbox{1.5cm}{#1}\hfill
  \parbox{12cm}{\beanon #2\eeanon}\hfill
  \parbox{1cm}{\bea\eea}\newline}
\ncm{\nr}[1]{\parbox{1cm}{\begin{eqnarrayabc}#1\end{eqnarrayabc}}\\}

\ncm{\kal}[1]{\mbox{$\cal #1 $}}
\ncm{\mrk}[1]{\!\!\! #1 \!\!\!} 
\ncm{\qed}{\hspace*{0.4cm}\rule{0.24cm}{0.24cm}}  
\ncm{\mbold}[1]{\mbox{\boldmath $ #1 $}}   
\ncm{\bm}{\mbold}
\ncm{\str}{\stackrel}
\ncm{\sub}{\subset}
\ncm{\e}{\varepsilon}
\ncm{\ka}{\kappa}
\ncm{\inputc}[1]{\begin{center}\input{#1}\end{center}}
\ncm{\lto}{\longrightarrow}
\ncm{\x}{\times}
\ncm{\bmm}{\bm{\cal M}}
\ncm{\cp}{{\bf P}}    
\ncm{\bfp}{{\bf P}}
\ncm{\bmi}{\bm{i}}
\ncm{\bmom}{\bm{\om}}
\ncm{\bmOm}{\bm{\Om}}
\ncm{\res}{\restriction}
\ncm{\bmL}{\bm{\cal L}}
\ncm{\bmell}{\bm{\ell}}
\ncm{\bmE}{\bm{\cal E}}
\ncm{\bme}{\bm{e}}
\ncm{\bmpi}{\bm{\pi}}
\ncm{\bmr}{\bm{r}}
\ncm{\bmsigma}{\bm{\sigma}}
\ncm{\wt}{\widetilde}
\newcommand{\beaa}{\begin{eqnarray}}
\newcommand{\eeaa}{\end{eqnarray}}

\begin{document}

\author{{\bf Oscar Bolina}\thanks{Supported by FAPESP under grant
97/14430-2. {\bf E-mail:} bolina@math.ucdavis.edu} \\
%EndAName
Department of Mathematics\\
University of California, Davis\\
Davis, CA 95616-8633 USA\\
}
\title{\vspace{-1in}
{\bf Difficult Problems Having Easy Solutions}}
\date{}
\maketitle
\begin{abstract} 
\noindent
{\it We discuss how a class of difficult kinematic problems can 
play an important role in an introductory course in stimulating 
students' reasoning on more complex physical situations. The 
problems presented here have an elementary analysis once certain 
symmetry features of the motion are revealed. We also explore 
some unexpected directions these problems lead us.} 
\newline
\noindent
{\bf Key Words:} Plane Motion, Central Force Motion \hfill \break
\end{abstract}
\noindent

%%%%%%%%%%%%%%%%%%%%%%%%%%%%%%%%%%%%%%%%%%%%%%%%%%%%%%%%%%%%%%%%%%%%%%%%%%%%%%
%1
%%%%%%%%%%%%%%%%%%%%%%%%%%%%%%%%%%%%%%%%%%%%%%%%%%%%%%%%%%%%%%%%%%%%%%%%%%%%%%

\noindent
There is a class of kinematical problems whose general analysis 
lies beyond the level of undergraduate students but which nonetheless 
can play an important role in introductory courses. This happens 
because the problems involve certain symmetry features that allow for 
an easy solution once the symmetry is revealed.
\newline
What makes these problems worth mentioning here is that they
are particularly useful in stimulating students' reasoning on
richer physical situations. 
\newline
Consider this typical example. 
\vskip .1 cm
\noindent
{\normalsize \it A boat crosses a river of width {\it l} with 
velocity of constant magnitude {\sl u} always aimed toward a 
point {\sl S} on the opposite shore directly across its starting 
position. If the rivers also runs with uniform velocity {\sl u}, 
how far downstream from S does the boat reach the opposite shore?} 
\vskip .1 cm
\noindent
Fig. 1 depicts the situation when the boat is at point {\it B} in 
its path {\it RV} toward the opposite shore. Its velocity vector 
along {\it BS} makes an angle $\theta$ with {\it RS}. The river 
velocity is represented along {\it TB}.
\newline
The {\it (easy)} solution comes from the observation that 
in a time interval $\Delta t$ the distance {\it BS} of the 
boat to the point {\it S} on the opposite shore diminishes 
by $(u-u\cos\theta)\Delta t$, while, at the same time, its
distance {\it TB} to {\it T} increases by the same amount, 
thus keeping the sum $BS+TB$ constant throughout the trip 
along {\it RV}. Since it is clear that at the starting 
position $BS=l$ and $TB=0$, while at the final position 
$BS=BT=d$, which is the distance downstream, we conclude 
that $d=l/2$ (See \cite{R} for further examples). 
\newline
The importance of this example lies in its far-reaching scope. For
instance, many problems in central force motion (planetary motion) 
can be formulated in similar terms, and the same kind of analysis 
applies as well. Note the similarity we obtain if we turn the boat 
of the example into a comet describing a parabolic orbit around a 
sun {\it S} in the focus, as in Fig. 2. In this case, it is the 
component velocities along {\it TB} and normal to {\it SB} that 
are constants throughout the orbit (Compare with Fig. 1). As a 
result we learn that now the sum of the distances $SB+BP$ is 
constant along the comet's orbit. This fact defines the parabolic
orbit and, with $SB=r$ and $BP=r \cos\theta$, yields its equation 
in polar coordinates as $r(1+\cos\theta)={\rm const.}$
\newline
We can go still further to realize that a parabolic orbit is 
unstable under the influence of possible nearby planets, and may 
be easily converted into an ellipse or an hyperbola, according to 
whether the velocity of the comet decreases or increases as a 
result of the planetary perturbation \cite{A}. When this happens, 
the pattern of constant velocity components, {\it u} parallel to 
a fixed direction (as {\it BT}), and {\it v} normal to the 
radius vector {\it BS}, is preserved in the new orbit. The
difference is that these two velocities will not be equal to 
each other any more, and the ratio $u/v$ will determine the shape 
of the orbit (though not its size), being an ellipse when 
$u/v <1$, and a hyperbola when $u/v >1$.
\newline
We leave it to the reader to explore other features of
planetary motion opened up by our example.

%%%%%%%%%%%%%%%%%%%%%%%%%%%%%%%%%%%%%%%%%%%%%%%%%%%%%%%%%%%%%%%%%%%%%%%%%%%%%
%BIBL
%%%%%%%%%%%%%%%%%%%%%%%%%%%%%%%%%%%%%%%%%%%%%%%%%%%%%%%%%%%%%%%%%%%%%%%%%%%%%

\newpage
\begin{figure}
\centerline{
\epsfbox{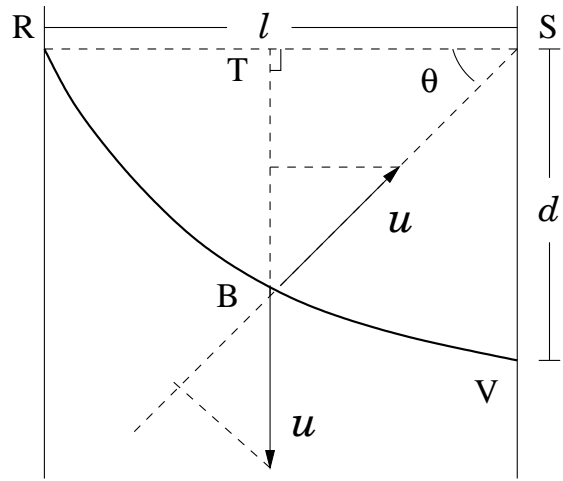}}
\caption{In this motion $SB+TB$ is constant.}
\end{figure}

\newpage
\begin{figure}
\centerline{
\epsfbox{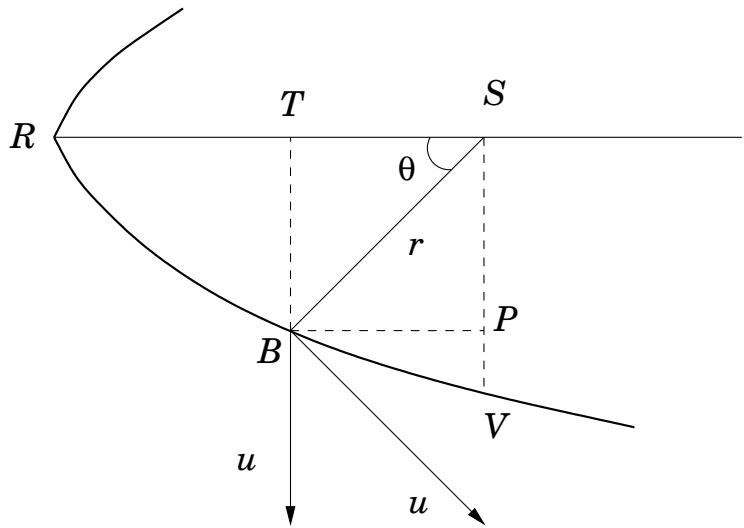}}
\caption{For the parabola, $SB+BP$ is constant.}
\end{figure}


\begin{thebibliography}{9}

\bibitem{R} A.S. Ramsey, {\it Dynamics}, Cambridge University
Press (1951) chapter 3. 

\bibitem{A} {\it The New Space Encyclopaedia}, E. P. Dutton 
$\&$ Co., INC. N.Y. (1960) Entry: Comet.

\end{thebibliography}
\end{document}